\documentstyle[12pt]{article}

\newcommand{\EQ}{\begin{equation}}
\newcommand{\EN}{\end{equation}}
\begin{document}
\topmargin 0pt
\oddsidemargin=-0.4truecm
\evensidemargin=-0.4truecm
\renewcommand{\thefootnote}{\fnsymbol{footnote}}
\newpage
\setcounter{page}{1}
\begin{titlepage}
\begin{flushright}
IC/92/432 \\
UMDHEP 93-105 \\
LMU-16/92 \\
December 1992
\end{flushright}
\vglue 0.4truecm
\begin{center}
{\large PLANCK-SCALE PHYSICS AND SOLUTIONS TO THE STRONG CP-PROBLEM
WITHOUT AXION }
\vspace{0.3cm}

{\large Zurab G. Berezhiani}\footnote{Alexander von Humboldt
Fellow}\footnote{E-mail: zurab@hep.physik.uni-muenchen.de,
{}~39967::berezhiani},\\
\vspace{0.2cm}
{\em Sektion Physik der Universit\"{a}t M\"{u}nchen, D-8000 Munich-2,
Germany\\
Institute of Physics, Georgian Academy of Sciences, Tbilisi
380077, Georgia\\}
\vspace{0.4cm}
{\large Rabindra N. Mohapatra}\footnote{E-mail: rmohapatra@umdhep,
{}~47314::rmohapatra } \\
\vspace{0.2cm}
{\em Department of Physics, University of Maryland, College Park,
MD 20742, USA } \\
\vspace{0.3cm}
and \\
\vspace{0.3cm}
{\large Goran Senjanovi\'{c}}\footnote{E-mail: goran@itsictp.bitnet,
{}~vxicp1::gorans} \\
\vspace{0.2cm}
{\em International Centre for Theoretical Physics,
I-34100 Trieste, Italy } \\

\end{center}
\begin{abstract}
We analyse the impact of quantum gravity on the possible solutions to
the strong CP problem which utilize the spontaneously broken discrete
symmetries, such as parity and time reversal invariance. We find that
the stability of the solution under Planck scale effects provides an
upper limit on the scale $\Lambda$ of relevant symmetry breaking. This
result is model dependent and the bound is most restrictive for the seesaw
type models of fermion masses, with $\Lambda < 10^6$ GeV.
\end{abstract}
\end{titlepage}
\renewcommand{\thefootnote}{\arabic{footnote}}
\setcounter{footnote}{0}
\newpage
{\bf 1. Introduction.} It is well known that the instanton effects bring
about a periodic structure of QCD vacuum. This leads to
$CP$-violation by strong interaction \cite{Kim}, characterized by the
$\bar{\Theta}$ parameter defined as
\EQ
\bar{\Theta}=\Theta_{QCD} + \Theta_{QFD}
\EN
Here $\Theta_{QCD}$ is the coefficient of the $P$- and $CP$-violating
gluonic anomaly term $G\tilde{G}$ and
$\Theta_{QFD}=$argDet$\hat{m}_u\hat{m}_d$ is a fermionic contribution, where
$\hat{m}_u$ and $\hat{m}_d$ are the up and down quark mass matrices. This
$CP$-violation manifests itself in an appearance of the neutron dipole
electric moment, which is known experimentally \cite{Smith} to be less than
$10^{-25}$ e cm leading to a phenomenological upper bound on $\bar{\Theta}$
of about $10^{-9}$ \cite{Crewther}. However, in the standard model
$\bar{\Theta}$ receives an infinite renormalization even if it is put to
zero at tree level by hand \cite{EG}. Understanding the smallness
of the $\bar{\Theta}$ without fine tuning of parameters is known as
the strong $CP$-problem.  There are two widely discussed approaches to
solving this problem:

i) The Peccei-Quinn mechanism \cite{Peccei} where the whole Lagrangian
of QCD plus QFD is required to obey a global $U(1)_{PQ}$ invariance with
nonvanishing color anomaly, which dynamically fixes $\bar{\Theta}=0$.
The spontaneous breaking of this symmetry leads to the
existence of a pseudo-Goldstone boson - axion \cite{WW,axion}.

ii) The discrete symmetry approach where a combination of discrete
symmetries such as $P$ or $CP$ is used to set $\bar{\Theta}=0$ naturally
\cite{Georgi,Segre,Nelson,Barr,BM,B,BCS} at the tree level. In such a
theory a finite $\bar{\Theta}$ arises at the higher loop level and
one has to show that $\bar{\Theta}<10^{-9}$.

An essential common ingredient of both these approaches is the
presence of global symmetries, either $U(1)_{PQ}$ or $P/CP$,
that guarantee the smallness of
$\bar{\Theta}$. However, these are not dynamical symmetries.
There is no physical ground to exclude that they are violated by higher
dimensional effective operators, that could
originate from new interactions existing at some high scale $M$.
These operators must be of non-renormalizable type so that at
$M\rightarrow \infty$ their effects disappear.
The ultimate scale for such higher order operators can be regarded as a
Planck scale $M_{Pl}$, where the gravity becomes
as strong as other interactions. This is a product of one's
experience with the quantum gravitational effects related to
virtual black holes \cite{Hawking} or wormholes \cite{Coleman} which
are likely not to respect global symmetries.
It is therefore important to include all higher
dimensional operators, consistent with the local invariance, in the
effective low energy theory before discussing whether the model solves
the strong $CP$-problem.\footnote{Some authors \cite{Nielsen} put
forward the idea that wormholes themselves may set $\bar{\Theta}$ to $0$
or $\pi$ dynamically, thereby avoiding the strong $CP$-problem. However,
as of now there is no universal agreement on the validity of
this point.} Since in most models the extra global symmetries
imposed on the Lagrangian are not automatic symmetries, one could
argue that perhaps the renormalizable terms in the low energy theory
should also be allowed to break these symmetries.\footnote{In a
self-consistent picture the global symmetry should originate as an
accidental symmetry of
the theory at lower energies, being automatically respected by the
renormalizable piece of the Lagrangian due to the certain field content.
The well-known examples are the lepton and baryon number conservation in
the standard model. The quantum gravity effects can violate them {\em only}
through the $d=5$ and $d=6$ operators  with $M\sim M_{Pl}$ \cite{BEG} - all
renormalizable terms are automatically invariant under these
symmetries. (In the context of grand unification these operators can appear
at the lower scale, with $M\sim M_{GUT}$ \cite{WeWi}.)
Neither $U(1)_{PQ}$ nor $P$ and $CP$ are automatic in general, though
there are a few attempts to introduce $U(1)_{PQ}$ as an accidental global
symmetry, at the price of enlarging the local symmetry of the theory
\cite{CPS}. It has been argued recently \cite{CKN},
that $P$ or $CP$ may also appear automatically as discrete {\em gauge}
symmetries in theories with dimensional compactification.}
In the absence
of detailed calculations of the non-perturbative quantum gravity effects
it is hard to argue for or against this. In this paper (as also in
refs. \cite{Kamion,GLR}) we will assume that  only the Planck scale induced
non-renormalizable terms are relevant. Certainly, they also have the
nice property of vanishing in the limit of zero gravity. Of course,
no such apology is needed if the theory is automatically invariant under
these global symmetries.

Such a study for the PQ
models was performed in recent papers \cite{Kamion}. It has been
shown that barring an unnaturally high degree of suppression of the
strength of the higher dimensional operators, the scale of $U(1)_{PQ}$
symmetry, $V_{PQ}$, must be less than 100 GeV in the simplest theories
of the invisible axion \cite{axion}, whereas the lower bound on
$V_{PQ}$ coming from various physical and astrophysical data is
larger by many orders of magnitude \cite{Kim}. This result considerably
diminishes our belief in Peccei-Quinn symmetry as a solution to the
strong $CP$-problem.\footnote{It is amusing to notice that the
original Peccei-Quinn model \cite{Peccei} with low scale axion
\cite{WW} is rather stable against Planck scale corrections. It is
however ruled out by experimental data.}

In this paper we consider the effects of gravity on the second
class of solutions. The original idea \cite{Georgi} was to use discrete
symmetries such as $P$ or $T$ in order to have $\Theta$-term vanishing
at tree level and to keep it finite and calculable in perturbation
theory. The challenge in this approach is to come out with
a simple enough model which gives $\bar{\Theta}$ sufficiently small.
The original models \cite{Georgi,Segre} suggested to
illustrate the philosophy behind them ended up using some {\em ad hoc}
further symmetries needed for the consistency of the program. One would not
have expected these symmetries to exist for any other
reason. Yet another difficulty of these models is that the Higgs
sector involved in electroweak symmetry breaking is nonminimal,
in which case the natural suppression of flavour-changing neutral
currents (FCNC) \cite{FCNC} does not occur. A more realistic scheme
was suggested by Nelson \cite{Nelson} and generalized by Barr
\cite{Barr}, which utilize $CP$ invariance to put $\Theta_{QCD}=0$, and
special field content to achieve also $\Theta_{QFD}=0$ at the tree
level after spontaneous breaking of $CP$. The key ingredients of these
models, which also avoid a problem of FCNC, are:

i) the presence of extra heavy fermions which are mixed with ordinary
quarks,

ii) the hypothesis that spontaneous $CP$-violation takes place only
in these mixing terms.

\noindent
The $\bar{\Theta}$-term effectively arises only at the 1-loop level.
It is less than $10^{-9}$ if the Yukawa coupling constants are
sufficiently small, less than $10^{-3}$.

However, the simplest possibility which utilizes the heavy fermions
came in paper \cite{BM} and subsequently in papers \cite{B,BCS}.
The idea of refs. \cite{BM,B} is based on the universal seesaw
mechanism \cite{seesaw,seesaw1}: the quark and lepton masses appear
due to their mixing with heavy fermions, in direct analogy with
the well-known seesaw picture for neutrinos \cite{SS}. In this picture
the solution to the strong $CP$-problem can be implemented through the
spontaneous violation of $P$-parity {\em only} \cite{BM}, as soon as one
deals with left-right symmetric model $SU(2)_L\otimes SU(2)_R\otimes
U(1)$. No other additional symmetry is required.
Alternatively, one can use a concept of $CP$-invariance (without
$P$-parity) even in the context of the $SU(2)\otimes U(1)$
model \cite{B}. This possibility, however, requires also
some extra symmetries (e.g. horizontal family symmetry, as also in the
model of Nelson \cite{Nelson}). One can show
that in  these models, with reasonable assumptions about the new scales
and parameters, the effective $\bar{\Theta}$ arising in loop effects
is small enough ($<10^{-9}$).

A different way to use the concept of parity was suggested in ref.
\cite{BCS}. The electroweak gauge symmetry of standard model
was doubled by introducing the mirror world with new weak interactions
being right-handed,
which repeats the whole pattern of
fermion masses of our left-handed world at some higher scale. This
is a result of spontaneous violation of $P$-parity between ordinary and
mirror worlds. The strong interactions are the same in both
sectors, so the contributions of mirror fermions cancel the infinite
renormalization of $\bar{\Theta}$ within the standard model and
$\bar{\Theta}$ is guaranteed to be negligibly small, less than $10^{-19}$.

In this paper we discuss the impact of the Planck scale effects on the
models of refs. \cite{BM}, \cite{B} and \cite{BCS}, which in the following
are referred to as BM, B and BCS models, respectively. We show that these
effects provide an upper bound on the scale of relevant symmetry breaking
($P$ or $CP$), with interesting phenomenological consequences.

{\bf 2. The BM model.} This model is based on the gauge
$SU(3)_c\otimes SU(2)_L\otimes SU(2)_R\otimes U(1)$ symmetry with the
quark fields in following representations:
\begin{eqnarray}
q_{Li}\hskip 0.2pc (1/2,0,1/3), \hskip 1pc U_{Ri}\hskip 0.2pc (0,0,4/3),
\hskip 1pc D_{Ri}\hskip 0.2pc (0,0,-2/3)\nonumber \\
q_{Ri}\hskip 0.2pc (0,1/2,1/3), \hskip 1pc U_{Li}\hskip 0.2pc (0,0,4/3),
\hskip 1pc D_{Li}\hskip 0.2pc (0,0,-2/3)
\end{eqnarray}
\noindent
where the $SU(2)_{L,R}$ isospins $I_{L,R}$ and $U(1)$ hypercharge $Y$ are
shown explicitly (the indices of the colour $SU(3)_c$ are omitted), and
$i=1,2,3$ is the family index. The Higgs sector consists of only two
doublets
\begin{eqnarray}
H_L\hskip 0.2pc (1/2,0,1) \nonumber \\
H_R\hskip 0.2pc (0,1/2,1)
\end{eqnarray}
Obviously, the fields in first rows of eqs. (2)-(3) have the usual
standard model content with respect to $SU(2)_L\otimes U(1)$ whereas
the fields in second rows form the analogous set of
$SU(2)_R\otimes U(1)$.\footnote{By adding the obvious lepton fields to the
quarks of eq. (2) the theory is free of gauge anomalies. As far as
strong CP-problem is concerned, we do not consider them here.}
The most general Yukawa couplings, consistent with gauge invariance,
are essentially the standard model ones:
\begin{eqnarray}
{\cal L}_L=\Gamma^{ij}_{Lu}\,\bar{q}_{Li}\,U_{Ri}\,\tilde{H}_L +
\Gamma^{ij}_{Ld}\,\bar{q}_{Li}\,D_{Ri}\, H_L + h.c.\nonumber \\
{\cal L}_R=\Gamma^{ij}_{Ru}\,\bar{q}_{Ri}\,U_{Li}\,\tilde{H}_R +
\Gamma^{ij}_{Rd}\,\bar{q}_{Ri}\,D_{Li}\, H_R + h.c.
\end{eqnarray}
For the singlet quarks $Q=U,D$ the mass terms
$\hat{M}_Q^{ij}\bar{Q}_{Li} Q_{Rj}$ are also allowed, unless
they are suppressed by some additional symmetry. Imposing the discrete
left-right symmetry $P_{LR}$, which is essentially parity \cite{LR}:
\EQ
q_L\leftrightarrow q_R,~~~~Q_L\leftrightarrow Q_R,~~~~
H_L\leftrightarrow H_R,~~~~W^{\mu}_L\leftrightarrow W^{\mu}_R
\EN
we have $\Gamma_{Lq}=\Gamma_{Rq}=\Gamma_q$ ($q=u,d$), and the mass
matrices $\hat{M}_Q$ are forced to be hermitean. The VEVs
$\,<H_L^0>=v_L\,$ and $\,<H_R^0>=v_R$, with $v_R\gg v_L=174$ GeV,
violate the $P_{LR}$ invariance and break the gauge symmetry down to
$U(1)_{em}$. As a result, the whole
$6\times 6$ mass matrices of quarks take the form
\begin{equation}
\begin{array}{cc}
 & {\begin{array}{cc} q_R & Q_R\end{array}}\\
{\cal M}~=~\begin{array}{c}
\bar{q}_L\\ \bar{Q}_L\end{array}&{\left(\begin{array}{cc}
0 & \Gamma v_L \\ \Gamma^{\dagger} v_R & \hat{M}
\end{array}\right)}\end{array}
\end{equation}
where $\Gamma =\Gamma_{u,d}$ and $\hat{M}=\hat{M}_{U,D}$ for the up- and
down-type quarks, respectively. Notice, that the
Det${\cal M}\sim$Det$\Gamma^{\dagger}\Gamma$ is real and therefore
$\Theta_{QFD}=0$. Since the $\Theta_{QCD}$ is absent from the beginning
due to parity invariance, we have $\bar{\Theta}=0$ naturally at tree
level. Then all that remains to do is to identify properly the fermion mass
eigenstates and show that the effective $\bar{\Theta}$ arising with
radiative corrections is sufficiently small. In fact,
the structure described above reflects the spirit of both BM and BCS
models, which are in fact two limiting cases, corresponding to
$\hat{M}\gg v_R$ and $\hat{M}\rightarrow 0$, respectively.

However, the quantum gravitational effects can induce the higher
dimensional operators violating explicitly the global $P_{LR}$ invariance
and thereby effectively contributing to $\bar{\Theta}$. These
operators should be cutoff by Planck scale $M_{Pl}$, so that their
effects disappear at $M_{Pl}\rightarrow \infty$. The leading order
terms allowed by gauge symmetry are the following:
\EQ
{\cal L}_5=\frac{1}{M_{Pl}}\,\bar{q}_{Li}\,(\alpha_u^{ij}\tilde{H}_L
\tilde{H}_R^{\dagger}+\alpha_d^{ij}H_L H_R^{\dagger})\,q_{Rj}\,+\,h.c. \\
\EN
\EQ
{\cal L}'_5=\frac{1}{M_{Pl}}\,\bar{Q}_{Li}Q_{Rj}\,(\beta_{RQ}^{ij}
H^{\dagger}_R H_R + \beta_{LQ}^{ij} H^{\dagger}_L H_L) + h.c. \\
\EN
\EQ
{\cal L}_6=\frac{1}{M_{Pl}^2}\,\bar{q}_{Li}\,(\gamma_{Lu}^{ij}U_{Rj}
\tilde{H}_L+\gamma_{Ld}^{ij}D_{Rj}H_L)\,H^{\dagger}_R H_R\,+
\,(L\leftrightarrow R)\,+\,h.c. \\
\EN
where the $\alpha,\beta$ and $\gamma$'s are in general the complex constants
of the order of one. Notice, that for these operators to be
$P$-invariant, the matrices $\alpha_q$ and $\beta_Q$ must be hermitean,
and $\gamma_{Lq}=\gamma_{Rq}^{\dagger}$.
Since we expect that the Planck scale effects are not to respect the
$P$-invariance, we assume the above matrices to be arbitrary.

Let us study now the impact of these operators on the $\bar{\Theta}$
parameter. It is convenient to assume that $\hat{M}\gg v_R$, in which
case the ordinary light quarks are essentially $q$'s,
whereas $Q$'s form a heavy states mixed with the latter through the
non-diagonal terms in eq. (6). The mass matrices of the $q$'s, induced
due to this, so called {\em universal seesaw}
mixing \cite{seesaw,seesaw1}, are the following:
\EQ
\hat{m}=v_Lv_R \Gamma \hat{M}^{-1} \Gamma^{\dagger}
\EN
The inter-family hierarchy (hierarchy between eigenvalues of $\hat{m}$)
can be related either with corresponding hierarchy in $\Gamma$'s or with
the {\em inverted} hierarchy \cite{BR} of the eigenvalues of $\hat{M}$'s.
As we have seen above, $\bar{\Theta}$ is vanishing at tree level.
It was shown in ref. \cite{BM} that a finite and small $\bar{\Theta}$
arises at the two loop level, whose magnitude can be less than $10^{-9}$
for the reasonable choice of parameters in the theory. However, the
Planck scale operator (7) will change the mass matrix (6) to the form:
\begin{equation}
\begin{array}{cc}
{\cal M}+\Delta{\cal M}~=~
&{\left(\begin{array}{cc}
\alpha v_Lv_R/M_{Pl}  & \Gamma v_L \\
\Gamma^{\dagger} v_R & \hat{M} \end{array}\right)}
\end{array}
\end{equation}
(Other contributions are neglected). Since the coefficients $\alpha$ are in
general complex, the effective $\bar{\Theta}$ is induced:
\EQ
\bar{\Theta}\simeq\frac{1}{M_{Pl}}\,{\rm Tr}(\alpha
\Gamma^{\dagger^{-1}}\hat{M}\Gamma^{-1})=
\frac{v_Lv_R}{M_{Pl}}\,{\rm Tr}(\alpha \hat{m}^{-1})
\EN
Obviously, the dominant contribution in eq. (12) comes from the light
quarks $u$ and $d$ with masses $\sim$few MeV. Then the condition
$\bar{\Theta}<10^{-9}$ constrains the scale of right-handed current $v_R$.
Demanding
that both the moduli and phases of the $\alpha$'s are $O(1)$ (certainly,
one should not exclude the possibility of an order of magnitude suppression),
and barring unforeseen conspiracies, we therefore conservatively estimate
an upper limit on $v_R$ of about $10^6$ GeV. As long as the seesaw formula
(10) is assumed to be valid, i.e. $\hat{M}\gg \Gamma v_R$, this limit
is rather independent of the details of the model. It equally applies
to the original version of BM model \cite{BM}, where the heavy $Q$ fermion
masses $M$ are assumed to be of the same order and the inter-family hierarchy
is related to the hierarchy of Yukawa couplings in eq.(4), as well as to the
inverse hierarchy model \cite{BR}, where all $\Gamma$'s are assumed to be
$O(1)$ and the inter-family hierarchy is originated from the hierarchy
in $\hat{M}$'s.

{\bf 3. The BCS model}. This model also utilizes the discrete $P_{LR}$
symmetry acting on the set of fermions as in eq. (2) and scalars as in eq.
(3). However, the mass terms $\hat{M}_Q$ of $Q$'s are put to zero due to
additional axial symmetry $U(1)_A$. The $U(1)_A$ hypercharges are defined as
following: $Y_A=Y$ for the fields of the first rows of eqs. (2) and (3),
and $Y_A=-Y$ for the second rows. It is obvious that incorporating this
symmetry, the theory remains free of gauge
anomalies. It can be local or global.\footnote{The original version
\cite{BCS} of the BCS model is based on the {\em local} symmetry
$SU(3)_c\otimes [SU(2)_L\otimes U(1)_L]\otimes [SU(2)_R\otimes U(1)_R]$,
where $SU(2)_L\otimes U(1)_L$ acting on the fields $q_L, Q_R$ and $H_L$
corresponds to the standard model of electroweak interactions and
$SU(2)_R\otimes U(1)_R$ with the fields $q_R, Q_L$ and $H_R$
corresponds to the parallel mirror world, a complete replice of ours,
but with new weak interactions being right-handed. These two worlds
communicate only via the same colour $SU(3)_c$. No doubt that apart
from nice "mirror" philosophy behind it, such a presentation is
completely equivalent to that we consider above: $U(1)_L\otimes U(1)_R=
U(1)\otimes U(1)_A$ with $Y=Y_L+Y_R$ and $Y_A=Y_L-Y_R$. Moreover,
in our case $U(1)_A$ can be global as well. It cannot serve us as a
Peccei-Quinn symmetry, being free of gauge anomalies.
As we show below, the impact of the Planck scale physics for the case
of global $U(1)_A$ is different from the case of the local one.}

Forbidding the explicit mass terms, the $U(1)_A$ symmetry has nothing
against the Yukawa couplings in eq. (4). As far as the fermion mass
spectrum and mixing is concerned, this model completely operates with
the parameters of the standard model.
Two fermion sectors are completely decoupled in the mass matrix (6):
$\hat{m}=\Gamma v_L$ is a mass matrix of the ordinary quarks $q_L$ and
$Q_R$, whereas the mass matrix of the mirror ones $q_R$ and $Q_L$
is just rescaled by the factor $v_R/v_L$.
At the tree level their contributions in $\bar{\Theta}$ cancel each other:
$\bar{\Theta}=$argDet$\Gamma^{\dagger}\Gamma=0$, and the
non-vanishing contribution to $\bar{\Theta}$ arising only at higher loops
is extremely small. Indeed, it was shown by Ellis and
Gaillard \cite{EG} that in standard model $\bar{\Theta}$, once put
to zero at tree level, arises only at the 3-loop level and
is about $10^{-19}$. The divergent contributions appear {\em only}
at the 6-loop level. However, in BCS scenario these are cancelled by
contributions of mirror quarks: $v_R$, as the scale of the mirror
(or parity) symmetry breaking provides the natural cutoff. This scale
can be arbitrarily large and so leaves us with a little hope of
detecting the mirror fermions.

The Planck scale operators, however, provide an upper bound on $v_R$.
Let us consider first the case of $U(1)_A$ symmetry being local,
as it was suggested in the original version of the BCS model.
In this case the effective $d=5$ operators of eq. (8) are forbidden
by local symmetry and the dominant contributions to $\bar{\Theta}$ come
from the $d=6$ operators of the eq. (9). One has:
\EQ
\bar{\Theta}\simeq \frac{v_R^2}{M_{Pl}^2}\,{\rm Tr}(\gamma\Gamma^{-1})
\EN
Then, by considering the contributions of the light quarks being
dominant, the condition $\bar{\Theta}<10^{-9}$ constrains $v_R$ to be
less than about $10^{12}-10^{13}$ GeV.

In the case of $U(1)_A$ symmetry being global both $d=5$ operators (7)
and (8) are active. Then the fermion mass matrices
take the form:
\begin{equation}
\begin{array}{cc}
{\cal M}+\Delta{\cal M}~=~
\left(\begin{array}{cc}
\alpha v_Lv_R/M_{Pl}  & \Gamma v_L \\ \Gamma^{\dagger} v_R &
\beta v_R^2/M_{Pl} \end{array}\right)
\end{array}
\end{equation}
so that the condition
\EQ
\bar{\Theta}\simeq\frac{v_R^2}{M_{Pl}^2}\,{\rm Tr}(\alpha
\Gamma^{{\dagger}^{-1}}\beta\Gamma^{-1})<10^{-9}
\EN
implies $v_R<10^9-10^{10}$ GeV.\footnote{Obviously, the same limit applies
to the general case of the model without $U(1)_A$ symmetry when the mass
terms of $Q$'s are allowed but are assumed to be less than $v_R$.} This
limit makes mirror world accessible at SSC/LHC, since in this case the
mirror partner of electron cannot be heavier than about $10$ TeV.

{\bf 4. The B model}. This model is also based on the field content of
eqs. (2)-(3), but in addition it utilizes also concept of local horizontal
symmetry $SU(3)_H$ \cite{SU3}: the fermions of the first row in eq. (2)
transform as triplets of $SU(3)_H$ and of the second row as anti-triplets,
while the scalars in eq. (3) are $SU(3)_H$ singlets. In fact, this is
exactly the field content of ref. \cite{seesaw} where the universal
seesaw mechanism was suggested for the generation of the quark and
charged lepton masses. Clearly, $SU(3)_H$ is free of gauge anomalies.
The matrices $\Gamma$ of the Yukawa coupling constants are forced
now to be $SU(3)_H$ singlets (i.e. proportional to the unit
$3\times 3$ matrix). The explicit mass terms $\hat{M}_Q$ are forbidden by
horizontal symmetry, but they appear due to Yukawa couplings
$G_{nQ}\bar{Q}_LQ_R\xi_n$, where $\xi_n$ ($n=1,2,..$), are some scalar
fields in representations $3$ and $\bar{6}$ of $SU(3)_H$, introduced for
the breaking of horizontal symmetry. Therefore, the mass matrices
of the heavy fermions $Q=U,D$ have the form:
\EQ
\hat{M}_{Q}=\sum G_{nQ}<\xi_n>
\EN
Provided that $\hat{M}_Q>v_R$, mass matrix of the ordinary quarks $q$
appears due to their seesaw mixing with $Q$'s. Since the Yukawa couplings
$\Gamma$ are the same for each family, the inter-family hierarchy between
$q$'s is necessarily related to the {\em inverse} hierarchy of the masses
of $Q$'s, which, on the other hand, reflects the hierarchy of the
horizontal symmetry breaking.

The presence of {\em chiral} horizontal symmetry $SU(3)_H$ makes it
unnatural to impose the left-right parity. However, CP-invariance can be
imposed, which implies that all the Yukawa couplings can be taken to be
real. The spontaneous $CP$-violation occurs in a sector of heavy fermions
due to relative phases of the VEVs $<\xi_n>$ \cite{B}, and is transfered
to the mass matrix $\hat{m}$ of $q$'s due to seesaw mechanism. However,
$\bar{\Theta}$ remains vanishing at tree level. It appears in radiative
corrections and can be rendered to be less than $10^{-9}$ under
certain assumptions on the parameters of the theory.

Let us include now the Planck scale effects. Instead of
$d=5$ operators (7), which are forbidden by horizontal symmetry, one
has to consider the $d=6$ operators
\EQ
\frac{1}{M_{Pl}^2}\,\bar{q}_{L}\,(\alpha_u^n\tilde{H}_L
\tilde{H}_R^{\dagger} + \alpha_d^{n}H_L H_R^{\dagger})\,
q_{R}\,\xi_n^{\dagger}\,+\,h.c. \\
\EN
Accounting for these operators, after the similar considerations as in
BM model one can deduce the limit $v_R<10^{10}$ GeV.
This constraint holds also true if instead of $SU(3)_H$ one considers the
left-right horizontal symmetry $SU(3)_{HL}\otimes SU(3)_{HR}$, with the
scalars $\xi_n$ being in representations $(\bar{3},3)$.
The $P$ parity is natural in this case, under which $\xi_n\leftrightarrow
\xi_n^{\dagger}$. Then the strong $CP$-problem can be solved
due to $P$-parity only, without imposing $CP$-invariance, since the
tree-level features of the model are essentially the same as in BM model.

On the other hand, if one deals with $CP$-invariance and $SU(3)_H$
symmetry, there is no need for $SU(2)_L\otimes SU(2)_R\otimes U(1)$
symmetry and the strong $CP$-problem can be solved at the level of
$SU(2)_L\otimes U(1)$ \cite{B}. The fermion content of the
theory with respect to $SU(2)_L\otimes U(1)\otimes SU(3)_H$ remains the
same apart from that now $q_R=u_R,d_R$ are electroweak singlets, as well as
$Q=U,D$. The scalar $H_R$ is also absent: the Yukawa couplings
in the second equation (4) are changed to the $SU(3)_H$-invariant mass
terms. Then the leading Planck scale operators consistent with the
$SU(2)_L\otimes U(1)\otimes SU(3)_H$ symmetry are the $d=5$ ones
\EQ
\frac{1}{M_{Pl}}\,\bar{q}_{L}\,(\alpha_u^nu_R\tilde{H}_L
 + \alpha_d^{n}d_RH_L)\,\xi_n^{\dagger} + h.c. \\
\EN
Repeating the above considerations concerning their contribution to
$\bar{\Theta}$, one can readily obtain the upper limit on the horizontal
symmetry breaking scale: $<\xi_n><10^6$ GeV, which is essentially also the
scale of $CP$-violation. This limit makes the flavour-changing
effects, related to the horizontal gauge bosons, available for the
experimental search in $CP$-violation phenomena or rare decays \cite{John}.
On the other hand, recalling
that in this model the hierarchy between ordinary quark and lepton families
should be the {\em inverse} with respect to the hierarchy in heavy
fermions, this bound simply means that the lightest among the latter are
expected to be in $100$ GeV to $1$ TeV range and so within the
reach of new accelerators.

{\bf 5. Conclusion.} We have shown that if the non-perturbative Planck
scale effects are assumed to break all global symmetries of nature,
viable mechanisms to solve the strong $CP$-problem can be constructed
implementing the natural physical symmetries such as parity or time
reversal invariance. The upper limit on the $\bar{\Theta}$
parameter imposes upper limits on the scale at which
these symmetries break. In two examples of such models \cite{BM,B},
this scale is less than $10^6$ GeV whereas in a third class of
models \cite{BCS}, the upper limit is $10^{12}$ to $10^{10}$ GeV. The
same consideration can be applied to the general class of models
\cite{Nelson,Barr}. In particular, in the original model of Nelson
\cite{Nelson} one can deduce the limit $<\xi><10^9$ GeV on the scale of
 horizontal symmetry breaking.

Many physical consequences can follow from these considerations. New
particles
and phenomena can be within the reach of new experiments, in particular,
at SSC/LHC. In models
\cite{BM,B} the obtained upper limits can be rephrased (in model dependent
way) in lower bounds on neutrino masses. On the other hands, the same
Planck scale effects could help us to avoide certain difficulties of the
models under consideration. Let us comment on two of them:

i) As is well known, the spontaneous breaking of discrete symmetries such as
$P$- and $CP$-parities leads to the formation of domain walls in the early
universe. It has recently been argued \cite{Balram}, that the same quantum
gravity effects can also induce the explicit $P$ and $CP$ violating Planck
scale non-renormalizable terms in the Higgs potential, that can cause the
decay of the domain walls and thereby avoide the associated cosmological
disaster. Of course, in the BCS type models the upper limit on the parity
breaking scale is high enough so that one could inflate away the
domain walls. The point is that the reheating temperature after inflation
must be less than the symmetry breaking scale so that the walls do not
reappear. As for the BM and B type models, one would require a low scale
inflation for the same purpose, which might  not be a best case for the
inflationary ideology.

ii) In the BCS type models there is another potential danger coming from
the fact that in the limit of exact $U(1)_A$ symmetry the lightest of heavy
mirror quarks (and leptons) is stable. Since one would expect that a baryon
asymmetry is associated with the mirror world as well, the cosmological
abundance of the latter should be much above the allowed experimental limit.
To inflate them away, one should remember that still low scale inflation
is needed: for example, in the case where the $U(1)_A$ symmetry is
global, their masses cannot be more than about $10^4-10^5$ GeV due to upper
limit $v_R<10^{10}$ GeV. However,  the Planck induced symmetry breaking
operators mix the ordinary and mirror
fermions (see eq. (14)), making the latter unstable and
thereby rendering them harmless. For the case of local $U(1)_A$ symmetry,
however, one again has to rely on inflation to achieve the same goal.
\vspace{0.3cm}

{\bf Acknowledgements.}

We thank E.Akhmedov, G.Dvali and B.Rai for discussions. The work of R.N.M.
is supported by a grant from the National Science Foundation and that of
Z.G.B. is supported by the Alexander von Humboldt Foundation.


\newpage
\begin{thebibliography}{99}
\bibitem{Kim} J.E.Kim, Phys.Rep. {\bf 150} (1987) 1; H.-Y.Cheng,
{\em ibid.} {\bf 158} (1988) 1; R.D.Peccei, in "CP-Violation",
ed. C.Jarlskog (World Scientific, Singapore, 1989) p.503.
\bibitem{Smith} K.Smith {\em et al.,} Phys.Lett. {\bf 234B} (1990) 191;
I.S.Altarev {\em et al., ibid.} {\bf 276B} (1992) 242.
\bibitem{Crewther} R.J.Crewther {\em et al.,} Phys.Lett. {\bf 88B}
(1979) 123.
\bibitem{EG} J.Ellis and M.K.Gaillard, Nucl.Phys. {\bf B 150} (1979)
141.
\bibitem{Peccei} R.D.Peccei and H.R.Quinn, Phys.Rev.Lett. {\bf 38}
(1977) 1440; Phys.Rev. {\bf D16} (1977) 1791.
\bibitem{WW} S.Weinberg, Phys.Rev.Lett. {\bf 40} (1978) 223;
F.Wilczek, {\em ibid.} 279.
\bibitem{axion} J.E.Kim, Phys.Rev.Lett. {\bf 43} (1979) 103;
M.Shifman, A.Vainshtein and V.Zakharov, Nucl.Phys. {\bf B 166} (1980) 493;
A.R.Zhitnitsky, Sov.J.Nucl.Phys. {\bf 31} (1980) 260;
M.Dine, W.Fischler and M.Srednicki, Phys.Lett. {\bf 104B} (1981) 199.
\bibitem{Georgi} H.Georgi, Hadronic J. {\bf 1} (1978) 155;
M.A.B.Beg and H.S.Tsao, Phys.Rev.Lett. {\bf 41} (1978) 278;
R.N.Mohapatra and G.Senjanovi\'{c}, Phys.Lett. {\bf 79B} (1978) 28.
\bibitem{Segre} G.Segre and H.Weldon, Phys.Rev.Lett. {\bf 42} (1979) 1191;
S.Barr and P.Langacker, {\em ibid.} 1654.
\bibitem{Nelson} A.Nelson, Phys.Lett. {\bf 136B} (1983) 387.
\bibitem{Barr} S.Barr, Phys.Rev.Lett. {\bf 53} (1984) 329;
Phys.Rev. {\bf D30} (1984) 1805.
\bibitem{BM} K.S.Babu and R.N.Mohapatra, Phys.Rev. {\bf D41} (1990) 1286.
\bibitem{B} Z.G.Berezhiani, Mod.Phys.Lett. {\bf A 6} (1991) 2437.
\bibitem{BCS} S.Barr, D.Chang and G.Senjanovi\'{c},
Phys.Rev.Lett. {\bf 67} (1991) 2765.
\bibitem{Hawking} S.W.Hawking, Comm.Math.Phys. {\bf 43} (1975) 199.
\bibitem{Coleman} G.Lavrelashvili, V.Rubakov and P.Tyniakov, JETP Lett.
{\bf 46} (1987) 167; S.W.Hawking, Phys.Lett. {\bf 195B} (1987) 337;
S.Giddings and A.Strominger, Nucl.Phys. {\bf B 307}
(1988) 854; S.Coleman, {\em ibid.} {\bf B 310} (1988) 643; S.J.Rey,
Phys.Rev. {\bf D39} (1989) 3185; B.Carter, in {\em General Relativity},
"An Einstein Centenary Survey", eds. S.Hawking and W.Israel, Cambridge
University Press, 1979.
\bibitem{Nielsen} H.B.Nielsen and M.Ninomiya, Phys.Rev.Lett. {\bf 62}
(1989) 1429; K.Choi and R.Holman, {\em ibid.} {\bf 62} (1989) 2575,
{\bf 64} (1990) 131; J.Preskill, S.Trivedi and M.B.Wise, Phys.Lett.
{\bf 223B} (1989) 26.
\bibitem{BEG} R.Barbieri, J.Ellis and M.K.Gaillard, Phys.Lett. {\bf 90B}
(1980) 249; E.Kh.Akhmedov, Z.G.Berezhiani and G.Senjanovi\'c, Phys.Rev.Lett.
{\bf 69} (1992) 3013.
\bibitem{WeWi} S.Weinberg, Phys.Rev.Lett. {\bf 43} (1979) 1566;
F.Wilczek, {\em ibid.} 1571.
\bibitem{CPS} J.E.Kim, Phys.Rev. {\bf D24} (1981) 3007;
H.Georgi, L.Hall and M.B.Wise, Nucl.Phys. {\bf B 192} (1981) 409;
P.H.Frampton, Phys.Rev. {\bf D25} (1982) 294; P.H.Frampton and T.Kephart,
{\em ibid.} 1459; K.Kang and I.Koh, {\em ibid.}
1700; K.Kang and S.Ouvry, {\em ibid.} {\bf D28} (1983) 2662;
D.Chang, P.B.Pal and G.Senjanovi\'c, Phys.Lett. {\bf 153B} (1985) 407;
D.Chang and G.Senjanovi\'c, {\em ibid.} {\bf 188B} (1987) 231;
Z.G.Berezhiani and M.Yu.Khlopov,
Yad.Fiz. {\bf 51} (1990) 1157 [Sov.J.Nucl.Phys. {\bf 51} (1990) 739];
Z.Phys. C {\bf 49} (1991) 73.
\bibitem{CKN} K.Choi,~D.B.Kaplan~and~A.E.Nelson,~Preprint~UCSD/PTH~92-11\
(1992);~{\em see~also}~\-
A.Strominger and E.Witten, Comm.Math.Phys. {\bf 101} (1985) 341;
C.Wetterich, Nucl.\-Phys. {\bf B 234} (1984) 413;
C.S.Lim, Phys.Lett. {\bf 256B} (1991) 233.
\bibitem{Kamion} R.Holman {\em et al.,} Phys.Lett. {\bf 282B} (1992)
132; M.Kamionkowski and J.March-Russel, {\em ibid.} 137; S.Barr and
D.Seckel, Phys.Rev. {\bf D46} (1992) 539; S.Ghigna, M.Lusignoli and
M.Roncadelli, Phys.Lett. {\bf 283B} (1992) 278.
\bibitem{GLR} D.Grasso,~M.Lusignoli~and~M.Roncadelli,~Phys.Lett.~{\bf 288B}\
(1992)~140;~E.Kh.Akh\-medov, Z.G.Berezhiani, R.N.Mohapatra and
G.Senjanovi\'c, Preprint IC/92/328, UMDHEP\- 93-020, LMU-12/92,
SISSA 149/92/EP  (1992) [Phys.Lett.B, in press].
\bibitem{FCNC} S.L.Glashow and S.Weinberg, Phys.Rev. {\bf D15} (1977)
1958; E.A.Paschos, {\em ibid.} 1964.
\bibitem{seesaw} Z.G.Berezhiani, Phys.Lett. {\bf 129B} (1983) 99.
\bibitem{seesaw1} Z.G.Berezhiani, Phys.Lett. {\bf 150B} (1985) 177;
Yad.Fiz. {\bf 42} (1985) 1309 [Sov.J.Nucl. Phys. {\bf 42} (1985) 825];
D.Chang and R.N.Mohapatra,
Phys.Rev.Lett. {\bf 58} (1987) 1600; A.Davidson and K.S.Wali,
{\em ibid.} {\bf 59} (1987) 393; S.Rajpoot, Phys.Lett. {\bf 191B} (1987)
122.
\bibitem{SS} M.Gell-Mann, P.Ramond and R.Slansky, in {\em Supergravity,}
eds. D.Freedman {\rm et al}., North-Holland, Amsterdam, 1979;
T.Yanagida, KEK Lectures, 1979 (unpublished); R.N.Mohapatra and
G.Senjanovi\'{c}, Phys.Rev.Lett. {\bf 44} (1980) 912.
\bibitem{LR} J.C.Pati and A.Salam, Phys.Rev. {\bf D10} (1974) 275;
R.N.Mohapatra and J.C.Pati, {\em ibid.} {\bf D12} (1975) 2558;
G.Senjanovi\'{c} and R.N.Mohapatra, {\em ibid.} 1502.
\bibitem{BR} Z.G.Berezhiani and R.Rattazzi, Phys.Lett. {\bf 279B} (1992)
124; Preprint LBL-32899, LMU-13/92 (1992).
\bibitem{SU3} J.L.Chkareuli, Pis'ma ZhETF {\bf 32} (1980) 684 [JETP Lett.
{\bf 32} (1980) 754].
\bibitem{John} Z.G.Berezhiani and J.L.Chkareuli, Yad.Fiz.{\bf 52} (1990)
601 [Sov.J.Nucl.Phys. {\bf 52} (1990) 383]; Z.G.Berezhiani, M.Yu.Khlopov
and R.R.Khomeriki, {\em ibid.} 538 [344].
\bibitem{Balram} B.Rai and G.Senjanovi\'{c}, Preprint IC/92/414 (1992).

\end{thebibliography}
\end{document}